%% file: main.tex
\documentclass[10pt, conference, letterpaper]{IEEEtran}
\IEEEoverridecommandlockouts

\usepackage{cite}
\usepackage{amsmath,amssymb,amsfonts}  
\usepackage{algorithmic}
\usepackage{graphicx}
\usepackage{textcomp}
\usepackage{xcolor}
\usepackage[compatibility=false]{caption}
\usepackage{subcaption}
\usepackage{xspace}
\usepackage{gensymb}
\usepackage{url}
\usepackage{hyperref}
\usepackage{booktabs} 
\usepackage{amsmath}  

\usepackage[T1]{fontenc}
\usepackage{newtxtext,newtxmath}
\usepackage{pifont}
\usepackage{enumitem}

\newcommand{\cmark}{\ding{51}}
\newcommand{\xmark}{\ding{55}}

\def\BibTeX{{\rm B\kern-.05em{\sc i\kern-.025em b}\kern-.08em
    T\kern-.1667em\lower.7ex\hbox{E}\kern-.125emX}}
    
\begin{document}

\title{Tiny-Twin: A CPU-Native Full-stack Digital Twin for NextG Cellular Networks}

\author{Ali Mamaghani$^{1}$, Ushasi Ghosh$^{1}$,  Srinivas Shakkottai$^{2}$ and Dinesh Bharadia$^{1}$, Ish Kumar Jain$^{3}$ \\
\normalsize{{$^{1}$University of California San Diego, CA, $^{2}$Texas A\&M University, TX}, $^{3}$Rensselaer Polytechnic Institute, NY,}\\
{\{amamaghani,\ ughosh,\ dineshb\}@ucsd.edu \quad \{sshakkot\}@tamu.edu} \quad \{jaini\}@rpi.edu}

\maketitle
\input{0_abstract}

\input{1_introduction}

\input{2_background}
\input{3_design}
\input{4_evaluations}

\input{6_conclusion}
\bibliographystyle{IEEEtran}
\bibliography{digitaltwin}

\end{document}

%% file: 0_abstract.tex
\begin{abstract}
Modern wireless applications demand testing environments that capture the full complexity of next-generation (NextG) cellular networks. While digital twins promise realistic emulation, existing solutions often compromise on physical-layer fidelity and scalability or depend on specialized hardware. We present Tiny-Twin, a CPU-Native, full-stack digital twin framework that enables realistic, repeatable 5G experimentation on commodity CPUs. Tiny-Twin integrates 
time-varying multi-tap convolution with a complete 5G protocol stack, supporting plug-and-play replay of diverse channel traces. Through a redesigned software architecture and system-level optimizations 
Tiny-Twin supports fine-grained convolution entirely in software. With built-in real-time RIC integration and per User Equipment (UE) channel isolation, it facilitates rigorous testing of network algorithms and protocol designs.
Our evaluation shows that Tiny-Twin scales to multiple concurrent UEs while preserving protocol timing and end-to-end behavior, delivering a practical middle ground between low-fidelity simulators and high-cost hardware emulators. We release Tiny-Twin as an open-source platform to enable accessible, high-fidelity experimentation for NextG cellular research.
\emph{Code available at:} \url{https://github.com/ucsdwcsng/Tiny_Twin}
\end{abstract}

%% file: 1_introduction.tex
\vspace{-0.2in}
\section{Introduction}
Next-generation cellular networks are evolving to support demanding applications like autonomous systems, extended reality (XR), and advanced IoT, each of them requires diverse application requirements beyond the traditional low-latency, reliability, and high-throughput wireless access. In response, the telecom industry is shifting toward highly programmable infrastructure such as Open Radio Access Network (Open-RAN), which allows for scenario-specific optimization through AI/ML-driven control loops.
This shift has sparked a new wave of wireless network co-design, where researchers develop AI-optimized 5G algorithms, such as link adaptation, mobility management, or RL-based resource allocation, that is tightly coupled with application requirements~\cite{edgeric}. However, understanding how these algorithmic decisions affect end-to-end performance remains a major challenge due to complex interactions across multiple layers of the RAN protocol stack. As a result, there is a growing need for highly available digital twins, which can accurately reflect the behavior of a real-world cellular network.

\begin{figure*}
\begin{center}
\includegraphics[width=0.95\linewidth]{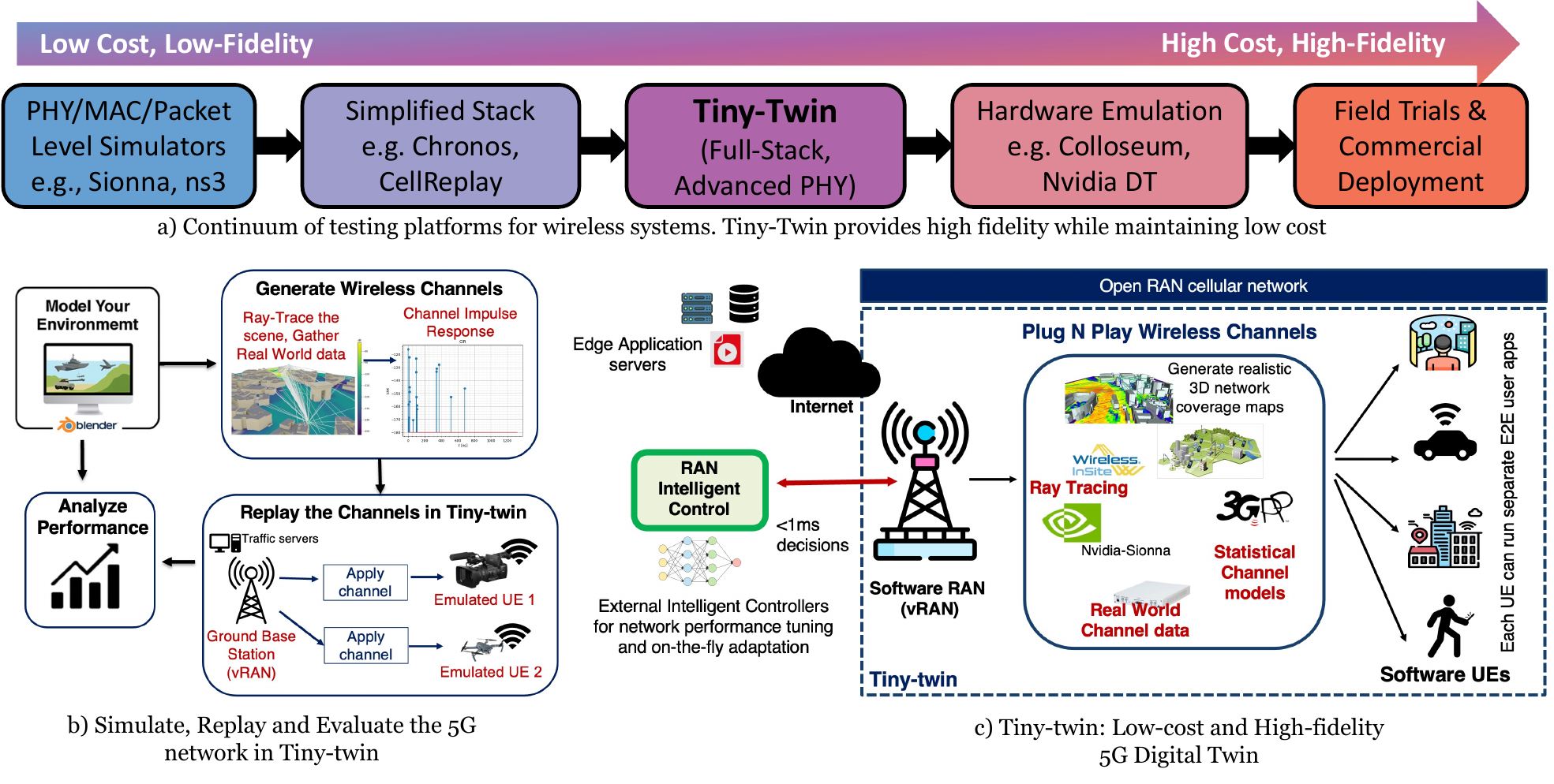}
\vspace{-0.1in}
\caption{Tiny-Twin Overview: Building a high-fidelity, low-cost   Digital Twin for Wireless Cellular Networks}
\label{fig:dt} 
\end{center}
\vspace{-0.3in}
\end{figure*}

\textbf{Such Digital Twin need to follow two main requirements: \textit{High-fidelity} and \textit{Low-cost}.}
High fidelity refers to the ability to emulate the full protocol stack with fine-grained physical layer attributes such as accurate channel models. Full-stack is required for capturing essential cross-layer effects such as retransmissions, buffering, and packet reordering. Without this realism, critical metrics like latency, jitter, and throughput become distorted, limiting the accuracy of the twin for evaluating AI-driven control loops and network co-design. 
It also requires accurate PHY channel modeling beyond simplified abstractions (e.g., averaged Signal-to-Noise Ratio SNR) or Channel Quality Indicator (CQI)
traces) to capture multipath fading, Doppler spread, and time-frequency selectivity-factors critical for realistic evaluation of functions like Modulation and Coding Scheme (MCS) selection\cite{adrx,ghosh2024sparc}, beamforming\cite{mimo-ric,jonnavithula2024beamarmor5g,zumegen2024beamarmor}, and scheduling\cite{radiosaber,seshasayee2026scout}.
The second requirement is a low cost, ensuring that such fidelity can be achieved on commodity CPUs rather than specialized FPGA- or GPU-based testbeds. This requirement is crucial for scalability and accessibility, enabling widespread experimentation and AI-optimized application development.

However, state-of-the-art Digital Twins do not meet these requirements as shown through three broad classes in Fig.\ref{fig:dt}(a).
The first class comprises basic low-cost simulators, such as MATLAB 5G toolbox, Sionna, or ns-3, with a focus on link-level behavior or simplified protocol models. While useful for targeted analysis, they lack integration with a full 5G protocol stack,  
limiting their utility for high-fidelity end-to-end evaluation. The second class primarily targets the upper layers of the protocol stack 
\cite{chronos, cellreplay}. While simplifying deployment, this abstraction also sacrifices fidelity, especially regarding PHY wireless channel impairments. The third class rely on custom hardware accelerators or FPGA-based platforms, needed to emulate the electromagnetic environment with high fidelity, such as Colosseum~\cite{colosseum} and Nvidia Digital Twin~\cite{nvidiatwin}. While accurate, these systems are high-cost, hardware-specific, and difficult to replicate with extended setup durations, forming a significant barrier to broader experimentation and rapid prototyping. Thus, there exists a trade-off between cost and fidelity. 

To address the critical need for
high-fidelity and low-cost wireless experimentation platform, we present \textbf{Tiny-Twin}—a CPU-Native, full-stack digital twin framework for 5G networks. Positioned as a middle ground between abstract L2 simulators and hardware-intensive emulators (Fig.\ref{fig:dt}a), Tiny-Twin is designed for a low-cost and high-fidelity framework. First, it delivers \textit{low-cost} by capturing fine-grained physical layer modeling entirely in software, allowing execution on commodity CPUs without reliance on specialized accelerators, thereby improving deployability and broadening accessibility. 

A key challenge is serving multiple computationally demanding User Equipments (UEs) in the digital twin environment 
This is possible by multiple system-level optimizations, such as parallelizing convolution computation, sparse convolution, and CPU pinning techniques (See Section~\ref{sec:tt_design}) Secondly, Tiny-Twin operates as a \textit{high-fidelity} plug-and-play channel replay engine (Fig.\ref{fig:dt} (b), (c)), uniquely capable of injecting time-varying channel traces into a live 5G full protocol stack built atop OpenAirInterface~\cite{oai}, achieving performance remarkably close to real time. Channel traces can originate from diverse sources, including 3GPP statistical models, ray-traced urban deployments, simplified synthetic abstractions, or, most importantly, complex real-world measurement campaigns. The ability to capture and replay real-world wireless environments with high fidelity is a defining strength of Tiny-Twin. It enables developers to emulate specific channel scenarios repeatedly and deterministically, supporting rigorous benchmarking, algorithm comparison, and application validation. 
Finally, its tight integration with a real-time RIC platform~\cite{edgeric} provides fine-grained control interfaces for AI-driven adaptation, enabling the development and debugging of closed-loop control policies that respond to sub-millisecond wireless fluctuations in a low-cost and high-fidelity environment.

We make the following contribution in this paper:
\begin{itemize}
    \item We design and implement Tiny-Twin, a low-cost, high-fidelity, CPU-native digital twin for full-stack 5G cellular networks, enabling realistic experimentation on commodity hardware without specialized accelerators.
    \item We identify key challenges in achieving channel realism and fast execution on CPUs and introduce system-level optimizations that enable real-time multi-tap channel convolution with over 20 channel taps.
    \item We demonstrate that Tiny-Twin scales to support up to 10 concurrent UEs while maintaining real-time operation and protocol-level fidelity.
    
    \item We develop an open-source Grafana-based monitoring framework that provides real-time visibility into PHY-, MAC-, and end-to-end system metrics, enabling detailed analysis and closed-loop experimentation.
    \item We make the Tiny-Twin codebase publicly available\footnote{Tiny-Twin software: \url{https://github.com/ucsdwcsng/Tiny_Twin}} to support reproducibility and facilitate further research in low-cost, high-fidelity cellular digital twins.
\end{itemize}

%% file: 2_background.tex
\section{Related Work}


\begin{table*}[htbp]
  \centering
  \small
  \renewcommand{\arraystretch}{0.9} 
  \caption{Comparison of Digital Twin frameworks}
  \label{comparison}
  \begin{tabular}{lcccccc}
    \toprule
    Framework 
      & \shortstack{Channel\\ Realism} 
      & \shortstack{Full-stack\\Processing} 
      & \shortstack{Support Protocol-\\level study} 
      & \shortstack{Hardware \\ Requirement} 
      & \shortstack{Compute \\ Time} 
      & \shortstack{Multi-user System \\ Scalability}\\
    \midrule
    Colosseum~DT \cite{dtcolosseum} 
      & \cmark & \cmark & \cmark & \textcolor{red}{FPGA} & 1ms & \cmark \\
    NVIDIA~DT    \cite{nvidiatwin}  
      & \cmark & \cmark & \cmark & \textcolor{red}{GPU} & 1ms & \cmark \\
    ns-3         \cite{ns-3}        
      & \xmark & \xmark & \cmark & CPU & \textcolor{red}{high} & \cmark \\
    Sionna \cite{sionna}        
      & \xmark & \xmark & \xmark & \textcolor{red}{GPU} & N/A & \xmark \\
    Chronos      \cite{chronos}     
      & \xmark & \cmark & \cmark & CPU & N/A & \cmark \\
    CellReplay   \cite{cellreplay}  
      & \xmark & \xmark & \xmark & CPU & N/A & \cmark \\
    Mahimahi   \cite{mahimahi}  
      & \xmark & \xmark & \xmark & CPU & N/A & \cmark \\
    OWDT   \cite{iye2025open}  
      & \cmark & \cmark & \cmark & \textcolor{red}{GPU} & \textcolor{red}{high} & \xmark \\
    OAI rfsim \cite{oai}  
      & \xmark & \cmark  & \cmark & CPU & \textcolor{red}{8ms} & \xmark \\
    SRSRAN ZMQ \cite{srs}  
      & \xmark & \cmark & \cmark & CPU & 1ms & \xmark \\
    UERANSIM \cite{ueransim}  
      & \xmark & \cmark & \cmark & CPU & N/A & \cmark \\

    \midrule
    Tiny-Twin (ours)                  
      & \cmark & \cmark & \cmark & \textbf{CPU} & \textbf{2ms} & \cmark \\
    \bottomrule
  \end{tabular}
\end{table*}

Table~\ref{comparison} highlights how our approach compares to existing cellular digital twin frameworks.

\textbf{High-Fidelity, High-Cost DT:} Frameworks that focus on high-fidelity emulation of the radio access network (RAN) include Colosseum’s digital twin platform~\cite{dtcolosseum}. It leverages USRP software-defined radios (SDRs) and FPGA-accelerated I/Q processing to emulate 5G signal propagation. 
This fidelity comes at a high cost, requiring terabytes of I/Q storage and commercial-grade SDR infrastructure.
Similarly, NVIDIA’s Aerial platform~\cite{aerial, nvidiatwin} utilizes A100 and H100 GPUs, as well as BlueField DPUs, to enable GPU-accelerated PHY simulation and ray-tracing-based wireless modeling. These platforms support research in AI-native RANs, but their reliance on datacenter-class hardware limits adoption by developers and researchers operating under resource constraints.
High-end GPU nodes (e.g., A100/H100) can cost \$25K–\$40K per server and require specialized cooling, whereas x86 CPU systems can be built for under \$3K, enabling scalable and cost-efficient deployment.
In contrast, Tiny-Twin is low-cost as it is computationally feasible on commodity CPU hardware, eliminating the need for FPGAs or GPUs. 
Tiny-Twin allows us to retain channel-aware features essential for protocol evaluation, while avoiding the full complexity of detailed PHY signal processing.

\textbf{Low-Fidelity Low-Cost  DT:} 
On the other hand, event-driven simulators such as ns-3~\cite{ns-3} and UERANSIM \cite{ueransim} have been widely adopted for 5G network research. But they often struggle to capture the full fidelity of a real-time 5G protocol stack. These platforms typically abstract away lower-layer timing constraints and complete protocol functionality, which contributes to the well-known sim-to-real gap. As a result, protocols and algorithms validated in such environments may fail to generalize or behave as expected under real-world deployment conditions.

In parallel, network emulators like Mahimahi~\cite{mahimahi} and CellReplay~\cite{cellreplay} 
are widely used in the systems community to evaluate transport protocols, adaptive bitrate (ABR) algorithms~\cite{pensieve}, and congestion control schemes~\cite{QUIC, ABC} by replaying pre-recorded network traces from commercial 
deployments. While convenient, these trace-driven tools lack a real protocol stack and do not capture well feedback-driven behaviors such as hybrid automatic repeat request (HARQ), scheduling adaptations, or stack-level retransmissions. They operate only at the network layer, offering no control over the 5G stack. As a result, they are less suitable for studying various cellular network function algorithms, such as adaptive bitrate policies, or application-aware scheduling, on end-to-end performance.
Other full-stack frameworks like Chronos~\cite{chronos} and \cite{tiny-twin} provide end-to-end 5G emulation environments, but they deliberately omit fine-grained PHY-layer modeling for scalability. 

\textbf{High-Fidelity Low-Cost  DT:} Full-stack digital twins trying to achieve high fidelity, such as the Open Wireless Digital Twin (OWDT)\cite{iye2025open}, focus on modeling a single UE within a simulated environment. 
However, they do not offer a scalable system design that supports multiple users while maintaining PHY fidelity, and they suffer from high computational cost in a multi-user system. In contrast, Tiny-Twin provides high-fidelity in a multi-user system by developing a finite-tap, time-domain channel convolution engine with CPU optimizations to support realistic PHY effects for all users.

%% file: 3_design.tex
\section{Tiny-Twin System Design}
\label{sec:Design}

This section highlights the fundamental design principles for a low-cost, high-fidelity digital twin, presents benchmarks revealing key bottlenecks, and introduces our system design and optimization, including parallelized, sparse Convolution and CPU Pinning to overcome them.
\begin{figure*}
\begin{center}
\includegraphics[width=0.9\linewidth]{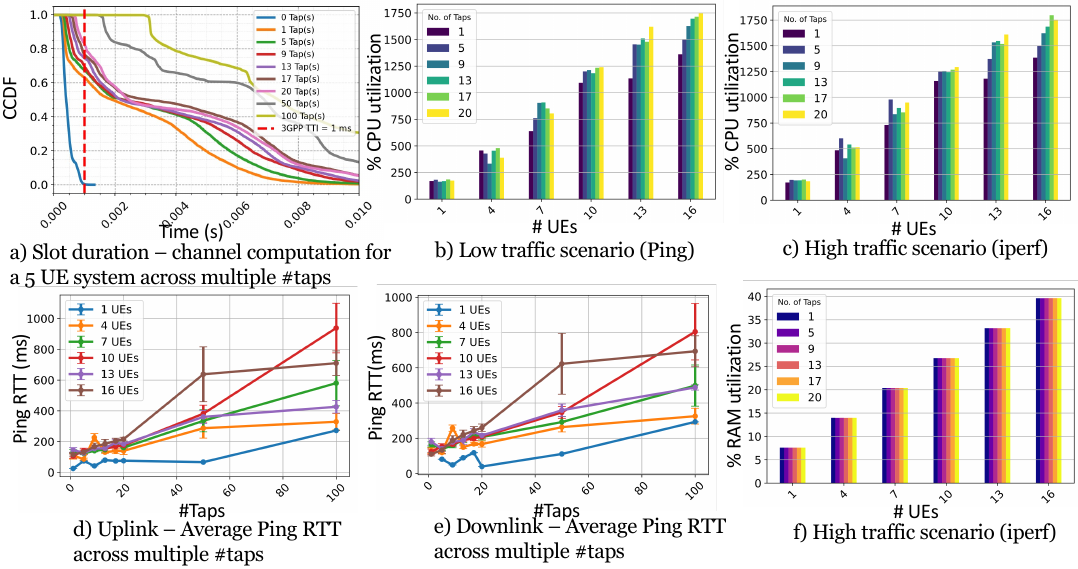}
\caption{Challenges with Vanilla system}
\label{fig:vanilla_challenges} 
\end{center}
\vspace{-0.3in}
\end{figure*}



\subsection{High-Fidelity, Low-Cost Digital Twin Design Principles}
\label{sec:design_principles}

Designing a practical digital twin for cellular networks requires balancing two competing objectives: \emph{high fidelity}, to faithfully reproduce real-world network behavior, and \emph{low cost}, to ensure accessibility and scalability on commodity hardware. In this work, we distill a set of design principles that define high fidelity in the context of next-generation cellular digital twins and guide the architectural choices behind Tiny-Twin.

High fidelity in a cellular digital twin extends beyond matching average link-level metrics such as signal-to-noise ratio (SNR), channel quality indicator (CQI), or throughput. Instead, fidelity must be defined in terms of the twin’s ability to reproduce \emph{end-to-end behavior under closed-loop operation}, where feedback mechanisms across the physical (PHY), medium access control (MAC), radio link control (RLC), transport, and application layers interact dynamically. Below, we define the requirements for a high-fidelity digital twin as illustrated in Table~\ref{comparison}. 

\begin{itemize}[leftmargin=*]
    \item 
\textbf{Channel realism} is central to this objective because wireless channels directly drive feedback loops, including link adaptation, HARQ retransmissions, scheduling decisions, and congestion control. Simplified abstractions, such as replaying SNR or CQI traces, fail to capture these interactions and often lead to misleading conclusions. A high-fidelity digital twin must therefore operate on \emph{IQ-level signals}, allowing physical-layer impairments to naturally propagate through the full protocol stack.

\item \noindent \textbf{Full-Stack Visibility and Cross-Layer Effects:}
High-fidelity digital twins must expose the \emph{entire cellular protocol stack}, from baseband signal processing to application-layer performance. While many simulators focus on isolated layers, such separation obscures cross-layer effects that are critical for evaluating modern cellular systems, particularly AI-driven control loops and application-aware scheduling. 

\item \noindent \textbf{Protocol-Level Study} involves the detailed examination and analysis of the functional layers above the physical (PHY) layer, including MAC, RLC, and higher transport protocols, within the digital twin's full 5G stack. 


\item \noindent \textbf{Low-Cost Through CPU-Native Design:}
Tiny-Twin delivers high-fidelity emulation on commodity x86 CPUs without GPUs or FPGAs, lowering cost while improving accessibility, reproducibility, and scalability.

\item \noindent \textbf{Compute Time Fidelity} is the ability to execute cellular slots quickly and consistently in wall-clock time. This is essential for large-scale AI/ML-driven RAN experimentation, where slow or highly variable slot execution can make training impractical and distort learned policies.

\item \noindent \textbf{Multi-user System Scalability} is the maximum number of concurrent UEs the twin can support while preserving physical-layer fidelity. This is critical for studying inherently multi-user NextG problems such as scheduling, interference management, and mobility under realistic network load.
\end{itemize}
\begin{figure*}[h!]
\begin{center}
\includegraphics[width=0.9\linewidth]{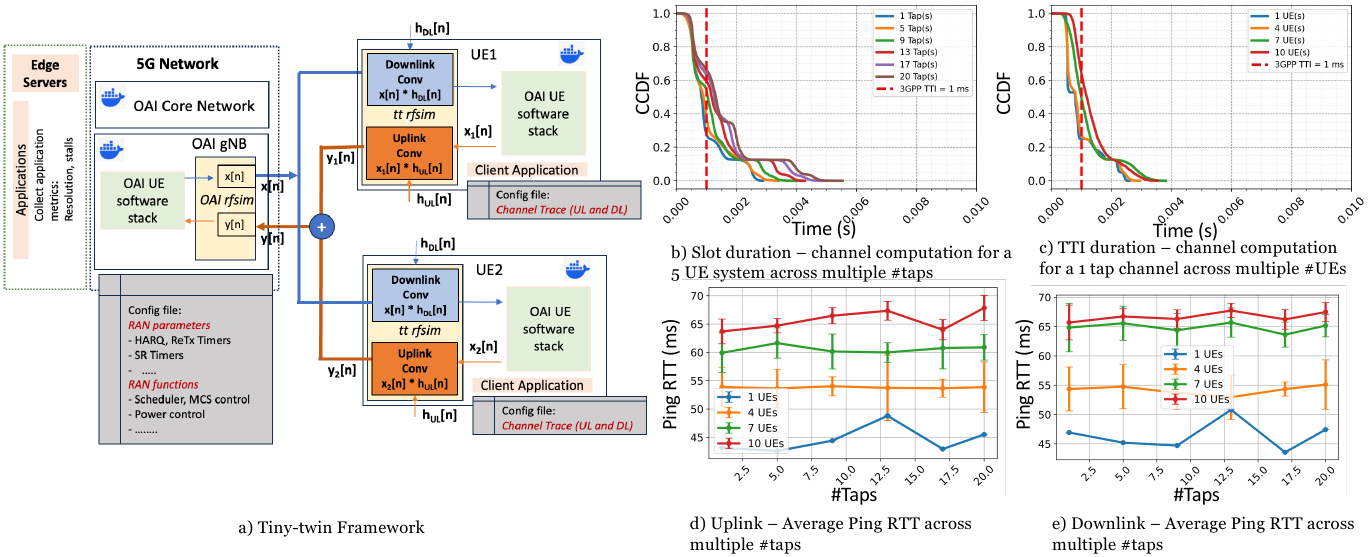}
\caption{Tiny-Twin system design and benchmarks}
\label{fig:tt-design} 
\end{center}
\vspace{-0.3in}
\end{figure*}

\subsection{System benchmarks and challenges with the vanilla system} 
We use the Open Air Interface (OAI) \cite{oai}, a widely adopted open-source 5G cellular stack, as the foundation for our framework. It supports a complete end-to-end 5G network, including the core, gNB, and UE components, executed in standalone containers, enabling modular deployment and controlled experimentation. 
OAI includes a built-in simulation mode called \texttt{rfsim}, which allows emulated RF signal exchange between the gNB and UEs over a TCP-based interface, it follows a telnet-style client-server model, where IQ samples are transmitted over socket connections between the gNB and each UE. In the default \texttt{rfsim} configuration, all channel convolutions are applied at the receiving node. This means, in downlink, the gNB transmits a raw (unconvolved) IQ stream over its server socket. Each UE acts as a telnet client, receives this stream, and applies a linear convolution with its own time-varying channel impulse response to simulate wireless propagation. In uplink, each UE transmits an unconvolved IQ stream to the gNB server. The gNB then performs channel convolution independently for each UE and aggregates the resulting streams. The channel effect is represented by: $ y_u(t) = x(t) * h_u(t) $, 
where \( x(t) \) is the transmitted IQ signal, \( h_u(t) \) is the UE-specific channel impulse response, and \( y_u(t) \) is the received, convolved signal. The convolution operator \( * \) denotes linear convolution
and varies independently for each receiver. However, in its default configuration, 
 the system applies only a single-tap scalar fading factor, effectively reducing the convolution to a per-symbol scaling operation, rather than a multi-tap filter. 
 We extended this functionality to support multi-tap channel convolution by enabling the \texttt{rfsim} module to read time-varying Channel Impulse Response (CIR) traces from external files, thus apply realistic, per-UE linear filtering. We refer to this setup as our vanilla system.


\label{sec:vanilla_benchmarks}
While the vanilla system provides a functional baseline for channel simulation, its architectural design leads to significant performance bottlenecks. As discussed earlier, all channel convolutions are performed at the receiver, which is particularly inefficient on the uplink. The gNB receives uplink IQ streams serially from each UE, applies convolution independently to each stream, and aggregates them only then. This serialized, compute-heavy pipeline introduces a critical bottleneck, especially under realistic multi-tap channel conditions. 


Figure \ref{fig:vanilla_challenges} presents a comprehensive benchmarking study of the vanilla system across multiple performance dimensions.  All experiments were performed on a system equipped with an AMD Ryzen 9 7900X processor (24 cores, 5.2 GHz), 32 GB RAM, and 1.8 TB of storage with a total system cost of approximately \$1528.00, ensuring a low cost compared to specific hardware. 
In Figure \ref{fig:vanilla_challenges}(a), we observe the TTI computation duration for a 5-UE setup under different channel tap configurations. We observe that the 90th percentile compute time reaches 8 ms even with a 10-tap channel. 
For larger tap values such as 50 or 100, the TTI durations rise significantly, clearly highlighting that convolution costs dominate processing time. These timing violations directly impact end-to-end performance, as seen in Figure \ref{fig:vanilla_challenges}(d) and \ref{fig:vanilla_challenges}(e), preventing the system from reaching high fidelity. Here, we observe that the uplink and downlink ping RTT increase sharply with tap count and number of UEs. Beyond 4 UEs and 10 taps, RTT often exceeds several hundred milliseconds, and connections may fail altogether.
Figure \ref{fig:vanilla_challenges}(b) and \ref{fig:vanilla_challenges}(c) show CPU utilization across low-traffic (ping) and high-traffic (iperf) scenarios, respectively. Utilization increases with the number of UEs, though it remains relatively stable across tap configurations, suggesting that user scaling is the dominant driver of compute load. Figure \ref{fig:vanilla_challenges}(f) shows RAM utilization under iperf, which also increases steadily with user count but remains within acceptable bounds. These results confirm that, despite not saturating CPU or memory resources, the system suffers from severe latency violations and degraded responsiveness due to inefficient compute allocation. TTI durations frequently exceed several ms, and ping round-trip times (RTTs) scale poorly with tap complexity and number of UEs. The core issue is that the PHY-layer channel convolutions are not effectively parallelized, resulting in underutilization of available system resources.
The subsequent subsections detail a restructured system that allows us to mitigate the identified bottlenecks and achieve high fidelity on a low-cost x86 CPU.

\subsection{Tiny-Twin Architecture Optimizations}
\label{sec:tt_design}
Building upon the insights from Section \ref{sec:vanilla_benchmarks}, we build the Tiny-Twin framework to efficiently emulate per-UE wireless scenarios, as illustrated in Figure \ref{fig:tt-design}(a). Built atop software-defined network components, Tiny-Twin enables end-to-end execution of real client applications and supports high-fidelity modeling of the signal chain down to the baseband IQ level.
Our system implementation involves the following optimizations:
\begin{itemize}[leftmargin=*]
    \item 
\textbf{Parallelizing Convolution Computation:} To address the primary bottleneck of the serial channel convolution at the gNB, as identified in Section \ref{sec:vanilla_benchmarks}, we adopt a UE-localized convolution processing approach. We update the OAI \emph{rfsim} mode to shift the entire channel convolution module into individual UE processes. As a result, both uplink and downlink channel convolutions are executed independently within each UE container, thus leveraging per-UE parallelism, distributing the most computationally intensive PHY operation across multiple processes and compute cores. 

\item\textbf{Sparse Convolution:} To reduce the computational cost per convolution operation, we employ a sparse convolution strategy. While each UE maintains the full multi-tap channel profile, convolution is performed only with the top-$n$ dominant taps at each time step. This significantly reduces the number of multiply-accumulate operations while preserving fidelity in modeling the channel's primary effects.

\item\noindent\textbf{CPU Pinning:} To reduce variability and scheduling overheads inherent in shared CPU environments, we employ CPU pinning. This dedicates specific logical CPU cores to key Tiny-Twin processes (e.g., two cores per UE container), reducing cache contention and improving processing consistency by reducing CPU switching time, particularly important for systems aiming for consistent timing behavior.

\end{itemize}
\textbf{Performance Benchmarks:}
We evaluate the performance of the Tiny-Twin system incorporating the architectural design and optimizations described above. Figure~\ref{fig:tt-design} (b) and (c) present the Transmission Time Interval (TTI) boundary distribution, demonstrating a significant reduction in latency compared to the unoptimized baseline (Figure \ref{fig:vanilla_challenges}a). The 90\textsuperscript{th} percentile TTI latency is reduced to approximately 2ms, representing a substantial 4x improvement in meeting timing constraints and achieving high fidelity.
Furthermore, Figures~\ref{fig:tt-design}(d) and (e) present the Ping RTT measurements across varying numbers of channel taps. Notably, the RTT remains largely constant regardless of the number of taps. 
These benchmarks demonstrate that the Tiny-Twin system can efficiently support detailed multi-tap channel convolution executed entirely on a standard CPU. Our design allows effective emulation with 20 taps per UE while maintaining performance metrics. This represents a highly fine-grained physical layer model achievable without requiring specialized hardware acceleration for the convolution task.

This finding challenges the conventional assumption that hardware acceleration is essential for PHY-intensive operations and establishes the feasible operational limits for channel detail 
when running a full-stack emulation environment on low-cost CPU-based compute platforms. 

%% file: 4_evaluations.tex
\section{Tiny-Twin System Evaluations}

To evaluate Tiny-Twin, we conduct a series of system-level experiments that span
channel integration pipelines, and full-stack responsiveness to dynamic wireless conditions. These evaluations validate our channel representation choices, demonstrate compatibility with diverse propagation environments, and highlight Tiny-Twin’s ability to capture end-to-end performance impacts often missed by abstract models.

\begin{figure}
\begin{center}
\includegraphics[width=\linewidth]{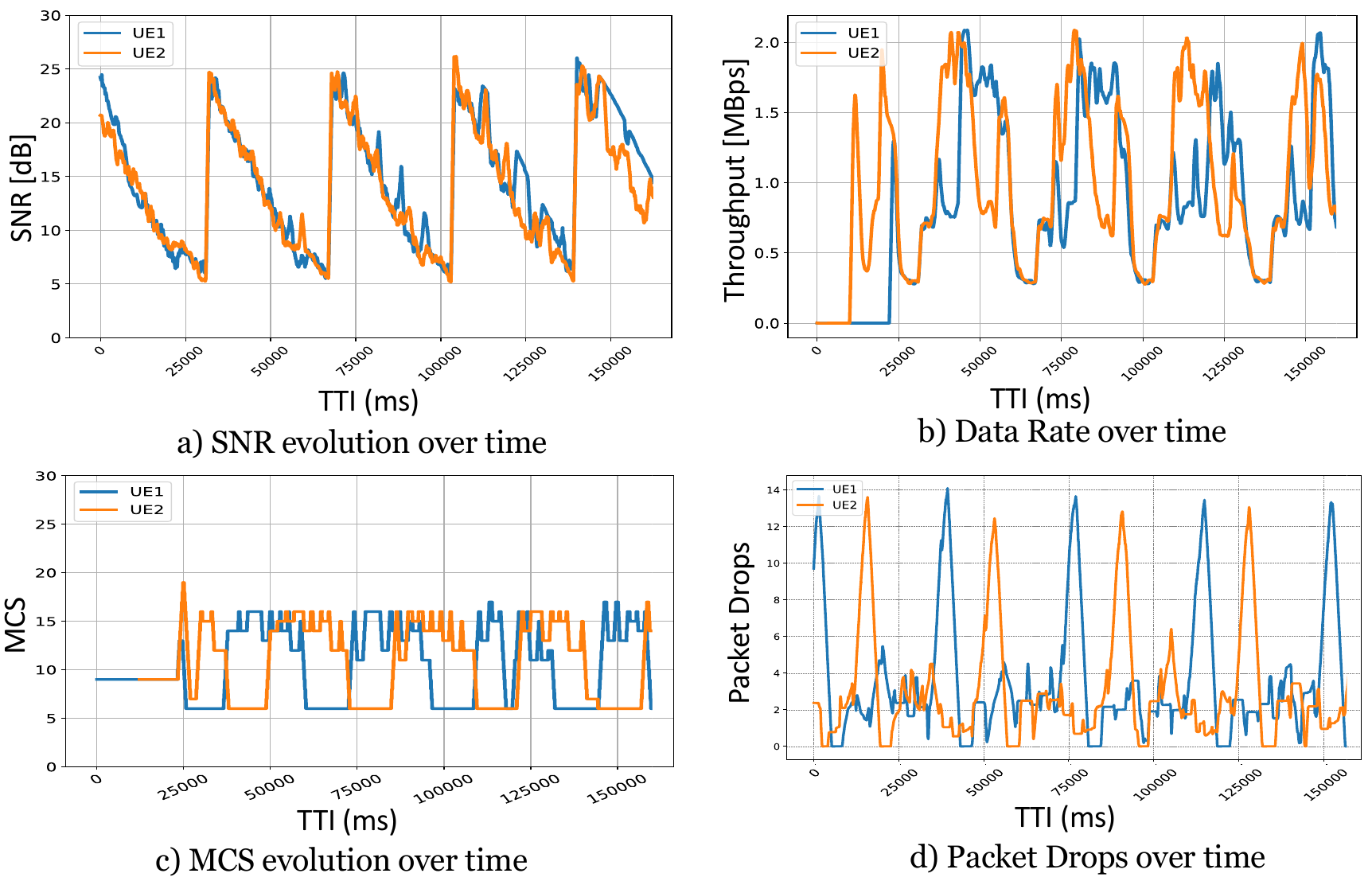}
\caption{Sanity Check of channel impelemetation on Tiny-Twin}
\label{fig:tt-microbenchmarks} 
\end{center}
\vspace{-0.3in}
\end{figure}

\begin{figure*}[!t]
\begin{center}
\includegraphics[width=0.9\linewidth]{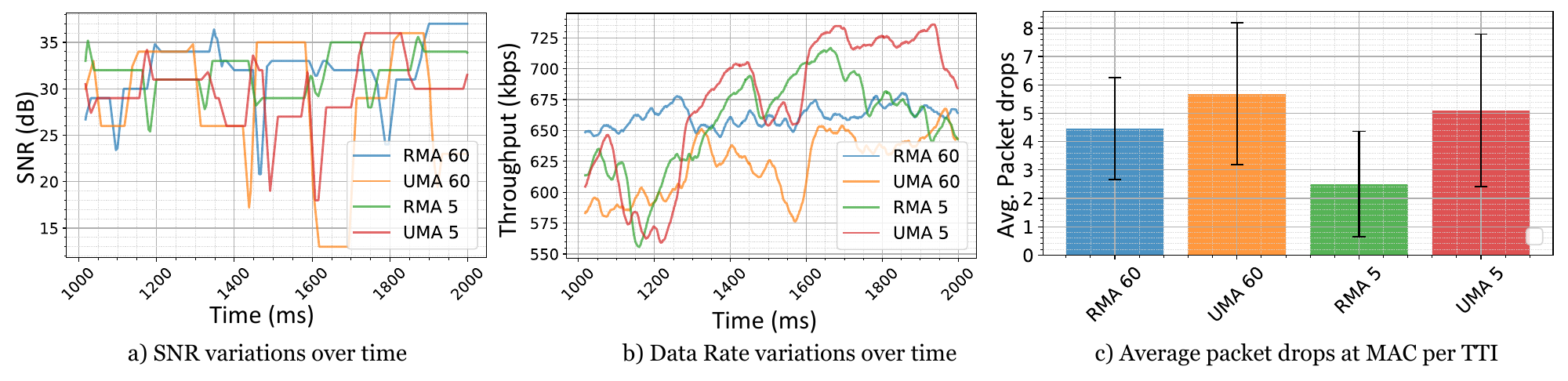}
\caption{Plug-N-Play Channels on Tiny-Twin}
\label{fig:3gpp_channel_and_metrics} 
\end{center}
\vspace{-0.2in}
\end{figure*}

\begin{figure*}[!t]
\vspace{-0.1in}
\begin{center}
\includegraphics[width=0.9\linewidth]{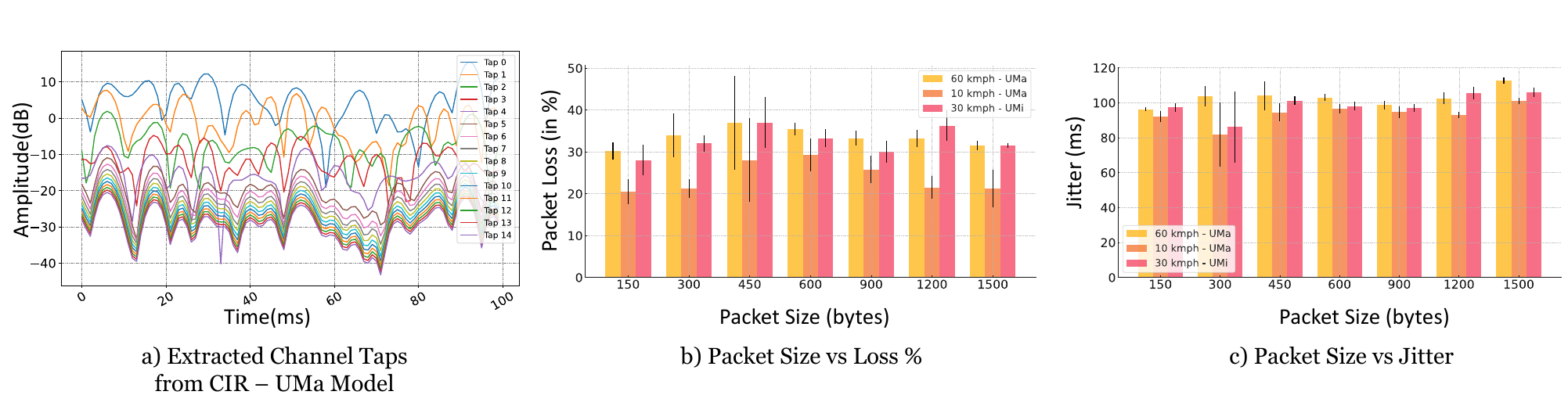}
\caption{E2E Application (IRTT) metrics on Tiny-Twin}
\label{fig:3gpp_channel_and_irtt} 
\end{center}
\vspace{-0.2in}
\end{figure*}
\noindent \textbf{Tiny-Twin compatible channel traces} Tiny-Twin enables experimentation with a wide range of wireless channels, including synthetic, ray-traced, model-based, or real-world over-the-air CIRs, by integrating them as time-varying, multi-tap CIR traces. These traces are represented as discrete complex tap values over time, designed to be fed into the per-UE channel convolution modules discussed in Section \ref{sec:Design}. To transform a given continuous CIR into this format, we perform a resampling process to extract uniformly spaced discrete channel taps onto a fixed delay grid for each time step. Crucially, these traces are structured with a temporal resolution of 1~ms, which is typically finer than the channel coherence time. This high-resolution sampling inherently captures the effects of Doppler spread and temporal fading dynamics, ensuring realistic time-varying behavior is reflected during emulation. Tiny-Twin thus provides a flexible framework for integrating diverse wireless conditions into a 5G network environment for end-to-end testing. 

In this study, we demonstrate Tiny-Twin’s flexibility by injecting three types of channel traces: (i) 3GPP statistical channels based on standardized PDPs for UMa, UMi, and RMa environments, with time variation synthesized via a Jakes-based Doppler model at 3.5 GHz \cite{ETSI_TR_138_901}; (ii) ray-traced channels generated using Sionna’s ray-tracing module \cite{sionna}, capturing site-specific multipath effects across indoor/outdoor LoS and NLoS scenarios by sampling CIRs along mobility paths; and (iii) real-world channels derived from the ARGOS dataset \cite{argos}. We will continue expanding our collection of channel traces, and upon acceptance, we will release all datasets along with the Tiny-Twin framework to support reproducible research.

\noindent \textbf{Sanity Check: Microbenchmarks with Synthetic channel trace on Tiny-Twin}
To verify the correctness of our channel implementation, we conduct a controlled microbenchmark using a synthetic channel trace where the SNR gradually degrades over time in a periodic fashion. This setup is designed to validate the convolution operations in the physical layer and their propagation through the stack. Figure~\ref{fig:tt-microbenchmarks} illustrates the evolution of key performance indicators, such as SNR, MCS, throughput, and packet drops, across two UEs under identical channel dynamics. As expected, the MCS adapts responsively to the changes in SNR, showcasing proper link adaptation. Correspondingly, we observe a drop in throughput as SNR decreases, and a rise in packet drops during low-SNR periods. Notably, the peaks in throughput and packet drops are inversely aligned, reaffirming that the system reacts to deteriorating channel conditions. Notably, these drops emerge from full-stack interactions that would be missed using simplified SNR or CQI-based traces. This demonstrates Tiny-Twin’s utility in capturing the nuanced effects of realistic channel dynamics on end-to-end 5G performance.
\begin{figure*}[!t]
\begin{center}
\includegraphics[width=0.95\linewidth]{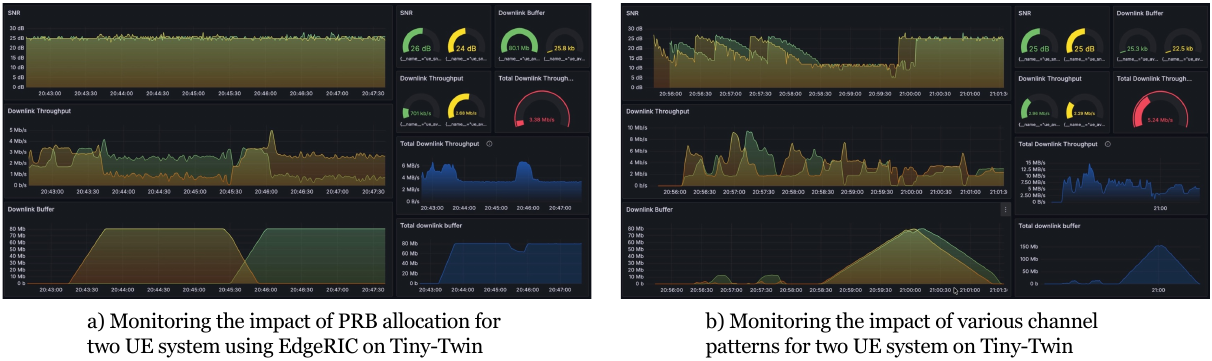}
\caption{The real-time monitoring graphical user-interface for Tiny-Twin}
\label{fig:monitoring} 
\end{center}
\vspace{-0.3in}
\end{figure*}

\noindent \textbf{Plug-N-Play Channels on Tiny-Twin}
Figure~\ref{fig:3gpp_channel_and_metrics} demonstrates Tiny-Twin’s ability to emulate diverse channel profiles across a range of mobility scenarios. Figure~\ref{fig:3gpp_channel_and_metrics}(a) shows that high-mobility channels (RMA/UMA 60) exhibit greater SNR fluctuations than their low-mobility counterparts. This variability directly impacts throughput Figure~\ref{fig:3gpp_channel_and_metrics}(b), where UMA 5 achieves the highest rates due to stable SNR, while RMA 60 shows pronounced dips. Figure~\ref{fig:3gpp_channel_and_metrics}(c) reveals that RMA 5 has the lowest MAC-layer packet drops, confirming the link between channel stability and reliability. 

Figure~\ref{fig:3gpp_channel_and_irtt} further illustrates how these PHY-layer dynamics translate into observable application-layer effects. Figure~\ref{fig:3gpp_channel_and_irtt}(a) presents CIR taps from a UMA model, revealing rich temporal multipath structure. Figure~\ref{fig:3gpp_channel_and_irtt}(b) and (c) show that increased mobility and larger packet sizes exacerbate packet loss and jitter, as captured using the IRTT tool~\cite{irtt}. Notably, the UMA 60 km/h trace consistently incurs higher jitter and loss than its lower-mobility counterpart. Together, these results demonstrate that Tiny-Twin preserves the end-to-end effects of physical channel variation, including effects often masked in simplified models that rely solely on SNR or CQI traces. 
\noindent \textbf{Monitoring framework on Tiny-Twin}
Figure~\ref{fig:monitoring} indicates Tiny-Twin’s monitoring framework for two different scenarios. We use Grafana  for the dashboard in combination with Prometheus  for data collection \cite{grafana,prometheus}. Figure~\ref{fig:monitoring}(a) shows the impact of asymmetric physical resource block (PRB) allocation on Tiny-Twin using EdgeRIC~\cite{edgeric}. Initially, EdgeRIC allocates a greater proportion of PRBs to the green UE and a smaller proportion to the yellow UE. This asymmetric resource allocation results in an observed higher throughput for the green UE and a corresponding larger buffer occupancy for the yellow UE. This allocation strategy is then reversed in the second phase, where the yellow UE is prioritized to quickly drain its accumulated buffer, while the green UE's throughput is reduced, causing its buffer to fill up. Figure~\ref{fig:monitoring}(b) specifies the impact of various channel patterns. Initially, the SNR has a repetitive pattern, and both the throughput and buffers follow the same pattern. Then the SNR drops, which leads to an upsurge in buffer. Finally, with good channel quality and a high SNR, the buffer size is reduced.

%% file: 6_conclusion.tex
\section{Conclusion and Future Work}

Tiny-Twin established a foundational utility by successfully demonstrating that high-fidelity, real-time Physical layer emulation is feasible on accessible commodity CPUs within a single-cell, multi-User environment. This foundational success provides a direct roadmap for expanding the system's operational scope, thereby maximizing the platform's research applicability and fully realizing the vision of a large-scale, accessible digital twin for NextG research. Our future efforts are focused on three core axes of architectural growth and research capability.

\subsubsection*{\textbf{Architectural Scalability}}


The single-CPU server evaluation proved the core innovation by enabling high-fidelity PHY emulation on commodity hardware, achieving fast compute for over 10 UEs.
Future development will rigorously implement distributed, cluster-based operation by leveraging the inherent isolation of UE processes into separate containers. This essential architectural evolution will eliminate the single-server bottleneck, allowing us to facilitate exhaustive large-scale (100+ user) simulations previously only possible on specialized, proprietary testbeds. 

\subsubsection*{\textbf{Extending Network Realism}}

To capture the complexity of real-world cellular deployments, our next objective is to expand the system's realism by integrating support for multi-base station operation and the systematic study of Inter-Cell Interference (ICI). This will directly leverage the planned cluster-based scaling.
Future work will extend the channel integration interface to programmatically allow each isolated UE process to receive and sum signals from all neighboring gNB cores across the cluster. This capability will unlock the rigorous testing of interference-sensitive network functions, including sophisticated advanced scheduling protocols under interference and critical mobility management (handover) protocols in a truly representative digital environment.




\noindent
\textbf{Acknowledgement:}
This work was funded in part by NSF
Grants CNS 2312978, CNS 1955696, ECCS 2030245, CNS
2312979 and ARO grant W911NF- 19-1-0367. 